\documentclass[%
 reprint,
 amsmath,amssymb,
 aps,
 nofootinbib,
 superscriptaddress
]{revtex4-2}

\usepackage{graphicx}
\usepackage{xcolor}
\usepackage{physics}
\usepackage{dcolumn}
\usepackage{bm}
\usepackage{braket}
\usepackage{amsmath}
\usepackage{amssymb}
\usepackage{enumerate}
\usepackage{booktabs}
\usepackage{txfonts}
\usepackage{listings}
\usepackage[subrefformat=parens]{subcaption}
\usepackage{caption}
\usepackage{multirow}
\allowdisplaybreaks[1]

\makeatletter
\def\hlinewd#1{%
	\noalign{\ifnum0=`}\fi\hrule \@height #1 %
	\futurelet\reserved@a\@xhline}
\makeatother

\begin{document}

\preprint{APS/123-QED}

\title{Quantum Speedup of Monte Carlo Integration with respect to the Number of Dimensions and its Application to Finance}

\author{Kazuya Kaneko}
\affiliation{Mizuho-DL Financial Technology Co., Ltd.\\ 2-4-1 Kojimachi, Chiyoda-ku, Tokyo, 102-0083, Japan}

\author{Koichi Miyamoto}
\email{koichi.miyamoto@qiqb.osaka-u.ac.jp}
\affiliation{Center for Quantum Information and Quantum Biology, Institute for Open and Transdisciplinary Research Initiatives, Osaka University \\ 1-3 Machikaneyama, Toyonaka, Osaka, 560-8531, Japan}
\affiliation{Mizuho-DL Financial Technology Co., Ltd.\\ 2-4-1 Kojimachi, Chiyoda-ku, Tokyo, 102-0083, Japan}

\author{Naoyuki Takeda}
\affiliation{Mizuho-DL Financial Technology Co., Ltd.\\ 2-4-1 Kojimachi, Chiyoda-ku, Tokyo, 102-0083, Japan}

\author{Kazuyoshi Yoshino}
\affiliation{Mizuho-DL Financial Technology Co., Ltd.\\ 2-4-1 Kojimachi, Chiyoda-ku, Tokyo, 102-0083, Japan}

\date{\today}

\begin{abstract}
Monte Carlo integration using quantum computers has been widely investigated, including applications to concrete problems.
It is known that quantum algorithms based on quantum amplitude estimation (QAE) can compute an integral with a smaller number of iterative calls of the quantum circuit which calculates the integrand, than classical methods call the integrand subroutine.
However, the issues about the iterative operations {\it in} the integrand circuit have not been discussed so much.
That is, in the high-dimensional integration, many random numbers are used for calculation of the integrand and in some cases similar calculations are repeated to obtain one sample value of the integrand.
In this paper, we point out that we can reduce the number of such repeated operations by a combination of the nested QAE and the use of pseudorandom numbers (PRNs), if the integrand has the separable form with respect to contributions from distinct random numbers.
The use of PRNs, which the authors originally proposed in the context of the quantum algorithm for Monte Carlo, is the key factor also in this paper, since it enables parallel computation of the separable terms in the integrand.
Furthermore, we pick up one use case of this method in finance, the credit portfolio risk measurement, and estimate to what extent the complexity is reduced.
\end{abstract}

\pacs{Valid PACS appear here}
                              
\maketitle

\section{\label{sec:intro}Introduction}

Monte Carlo integration is one of the important examples of computational tasks which quantum computers can speed up\cite{Montanaro,Suzuki}.
One of the reasons for its importance is the fact that it is widely used in industries, especially finance.
Financial firms are performing enormous Monte Carlo calculations for various purposes, so quantum speedup of such tasks may provide large impacts for them\footnote{See \cite{Hull} as a textbook of financial engineering and see \cite{Glasserman} as a reference which focuses on Monte Carlo methods used in finance}.
Some papers have already investigated how to apply the quantum algorithm for Monte Carlo to concrete problems in finance: for example, portfolio risk measurement\cite{Woerner,Egger,Miyamoto} and pricing of financial derivatives\cite{Rebentrost,Stamatopoulos,Martin,RamosCalderer,Vazquez,Kaneko}\footnote{See \cite{Orus,Egger2,Bouland} as reviews for application of quantum computing to finance, including Monte Carlo and other aspects.}.

The quantum algorithm for Monte Carlo integration is based on quantum amplitude estimation (QAE), which was originally investigated in \cite{Bassard} and also studied in the recent papers\cite{Suzuki,Aaronson,Grinko,Nakaji,Tanaka}.
It is often said that the quantum methods provide quadratic speedup compared with the classical method.
The meaning is as follows.
Both the quantum and classical Monte Carlo methods call the {\it oracle}, that is, the quantum circuit and the subroutine respectively, for calculation of the integrand.
In the former and the latter, the estimation error of the integral behaves as $O(N^{-1})$ and $O(N^{-1/2})$, respectively, where $N$ is the {\it oracle call number}.  
Equivalently, for the given tolerance $\delta$, the quantum and classical methods require the $O(\delta^{-1})$ and $O(\delta^{-2})$ oracle call, respectively.
Therefore, the quantum method can save the number of repeated oracle call tremendously.

On the other hand, in Monte Carlo integration, we often perform another type of repeated calculations, which has not been paid close attention to so far.
Specifically, when the dimension $D$ of the integration is very high, similar calculations can be repeated so many times in a call of the oracle, that is, in the flow for calculation of one sample value of the integrand.
Let us see a concrete example of this: credit portfolio risk measurement.
A credit portfolio is a collection of loans that a bank holds.
Each bank is monitoring some metrics which represent risks originating from defaults of obligors.
Major metrics include value at risk (VaR), the percentile point of the loss caused by defaults, and conditional VaR (CVaR), the expectation value of the loss under the condition that it is larger than VaR.
In calculating them using Monte Carlo, the values of the loss are randomly generated many times.
The flow of calculating a sample value of the loss is roughly as follows: (i) generate a random number (RN) $x$ for an obligor, (ii) determine whether he defaults or not according to $x$, (iii) if he defaults, add the exposure on him\footnote{This means the loss which arises if he defaults. In general, it is estimated by the product of the loan amount and the loss given default, the ratio of the amount which the bank fails to recover.} to the loss, then (iv) repeat steps (i)-(iii) for all obligors.
As this example shows, in the high-dimensional Monte Carlo integration where many RNs are necessary, we sometimes run many iterations of similar calculations, each of which uses a different RN.

In this paper, we propose a method based on QAE which speeds up such a type of repeated calculation.
In this new method, there are two key points to make QAE applicable.
First, it is necessary that the integrand is {\it separable}.
Although we will strictly state the meaning in Section \ref{sec:method}, the separable form roughly means that the contributions from different RNs to the integrand are separated into different terms.
This is necessary for computing the integrand separately for each dimension.
Second, this method uses pseudorandom numbers (PRNs).
PRN sequences are seemingly random but deterministic sequences generated by some recursion formulas.
In many cases, we can also use simple formulas to {\it jump} to the arbitrary position in the PRN sequence, that is, we can get the value of the $i$-th element not by repeatedly using its recursion formula $i$ times.
The authors originally proposed to use them in the quantum algorithm of Monte Carlo\cite{Miyamoto,Kaneko}.
In the case of the separable integrand, the use of a PRN sequence is crucial to achieve quantum parallel computation of separated terms.
Note that, when we use elements in a PRN sequence for a separable integrand, each element is used as a sample value of one of its arguments and thus determines a value of one of separated terms.
This implies that, we can construct a quantum circuit which receives an index specifying a term in the integrand as its input and gives a sample value of the term corresponding to the index.
Inputting a superposition of all indexes to this circuit, we can compute the terms in quantum parallelism.
Therefore, we can replace the naive iterative calculation with the QAE-based calculation, that is, a combination of quantum parallel computation of separated terms and summing them up by QAE.
The number of calling the circuit to a separated term changes from $O(D)$ to $O(\delta^{-1})$, which means that we can accomplish the reduction by a factor $O(\delta^{-1}/D)$.

However, we can not immediately conclude that the new method necessarily reduces computational time.
That is, time for calculating {\it one} term in the new method can be larger than that in the previous method.
This is because the new method replaces the recursive formula in the previous method with the jump formula and the latter is typically costly than the former.
If we write the times for calculating a term in the new and previous methods as $T_{\rm one,new}$ and $T_{\rm one,prev}$ respectively, computational time reduction by the new method is $T_{\rm one,new}\delta^{-1}/T_{\rm one,prev}D$.

Despite this point, we can find a concrete example where the new method actually reduces the total computational time.
We will take credit portfolio risk measurement as a concrete problem and the {\rm permutated congruential generator (PCG)}\cite{PCG}, which we originally proposed to use in the quantum algorithm for Monte Carlo in \cite{Miyamoto}, as a concrete PRN generator.
We will see that, in a typical setting, we can reduce the {\it T-count}, a popular metric of computational time cost defined later, by several tens of percent.

The rest of this paper is organized as follows.
In Section \ref{sec:rev}, we briefly review the quantum algorithm for Monte Carlo integration and use of PRN in it.
In Section \ref{sec:method}, we present the outline of the new method we propose.
In Section \ref{sec:example}, we consider application of the new method to credit portfolio risk measurement with PCG and estimate the expected speedup.
Section \ref{sec:summary} summarizes this paper.

\section{\label{sec:rev}The review of the quantum algorithm for Monte Carlo integration}

\subsection{The quantum algorithm for Monte Carlo integration}

Let us start with reviewing the quantum algorithm for Monte Carlo integration\cite{Montanaro}.
We here present the flow of calculating the expectation value $E[F(\vec{x})]$ of the function $F$ depending on $\vec{x}=(x_1,...,x_N)$, the vector of the $N$ stochastic variables.
It can be divided into the following four steps.
First, we create a superposition of possible values of $\vec{x}$ on a quantum register $R_{\rm RN}$ based on its probability distribution.
That is, we create $\sum_i \sqrt{p_i} \ket{\vec{x}_i}$, where $\vec{x}_i=(x_1^{(i)},...,x_N^{(i)}),i=1,2,...$ is the $i$-th possible value of $\vec{x}$, $p_i$ is the probability that $\vec{x}=\vec{x}_i$ and $\ket{\vec{x}_i}=\ket{x_1^{(i)}}...\ket{x_N^{(i)}}$ is the tensor product of states representing the values of the elements of $\vec{x}_i$.
Note that $x_j$ must be approximated in some discretized way if it is continuous.
Second, we calculate the integrand into another register $R_{\rm int}$ using $R_{\rm RN}$.
Note that the results for many patterns of $\vec{x}$ are simultaneously calculated in quantum parallelism.
Third, by controlled rotation, the integrand value is encoded into the amplitude of the ancilla qubit $R_{\rm ph}$.
Finally, using QAE \cite{Bassard,Suzuki,Aaronson,Grinko,Nakaji,Tanaka}, we estimate the probability that $R_{\rm ph}$ takes $\ket{1}$, which is equal to the expectation value we want.

From the first to the third steps, the quantum state is transformed as follows:
\begin{eqnarray}
& & \ket{0}\ket{0}\ket{0} \nonumber \\
& \rightarrow & \left(\sum_i{\sqrt{p_i}\ket{\vec{x}_i}}\right)\ket{0}\ket{0} \nonumber \\
& \rightarrow & \left(\sum_i{\sqrt{p_i}\ket{\vec{x}_i}}\ket{F(\vec{x}_i)}\right)\ket{0} \nonumber \\
& \rightarrow & \sum_i{\sqrt{p_i}\ket{\vec{x}_i}}\ket{F(\vec{x}_i)}\left(\sqrt{1-F(\vec{x}_i)}\ket{0}+\sqrt{F(\vec{x}_i)}\ket{1}\right) =: \ket{\Psi}.  \nonumber \\
& & \label{eq:QMC}
\end{eqnarray}
Here, the first, second and third kets correspond to $R_{\rm RN},R_{\rm int}$ and $R_{\rm ph}$, respectively.

We then explain the final step, QAE, based on \cite{Bassard}.
At first, we define some symbols.
We define $\theta$ as
\begin{equation} \ket{\Psi}=\cos(\theta\pi)\ket{\Psi_0}+\sin(\theta\pi)\ket{\Psi_1},0<\theta<\frac{1}{2},
\end{equation}
where $\ket{\Psi_0}$ and $\ket{\Psi_1}$ are the states where $R_{\rm ph}$ is $\ket{0}$ and $\ket{1}$ respectively.
Note that
\begin{equation} \sin^2(\theta\pi)=E_F:=\sum_i p_iF(\vec{x}_i)
\end{equation}
is the expectation value we want.
Besides, we write the operation corresponding to the whole of (\ref{eq:QMC}) as $A$ and define the operation $Q$ on the system consisting of $R_{\rm RN},R_{\rm int}$ and $R_{\rm ph}$ as
\begin{equation}
Q:=-AS_0A^{-1}S_1,
\end{equation}
where $S_0$ multiply the state by $-1$ if all qubits are $\ket{0}$ or do nothing otherwise and $S_1$ multiply the state by $-1$ if $R_{\rm ph}$ is $\ket{1}$ or do nothing otherwise. 
Then, preparing another register $R_{\theta}$ with $m$ qubits and using an algorithm containing $M-1$ iterations of calling $Q$, where $M=2^m$ and each $Q$ is controlled by $R_{\theta}$, we can create the state
\begin{equation}
\ket{\Phi_M(\theta)}:=\frac{1}{\sqrt{2}}\left(e^{i\theta\pi}\ket{\Psi_+}\ket{\phi_M(\theta)}-e^{-i\theta\pi}\ket{\Psi_-}\ket{\phi_M(1-\theta)}\right). \label{eq:QAEResult}
\end{equation}
Here, the second kets $\ket{\phi_M(\theta)}$ and $\ket{\phi_M(1-\theta)}$ correspond to $R_{\theta}$ and $\ket{\Psi_\pm}:=\frac{1}{\sqrt{2}}\left(\ket{\Psi_1}\pm i\ket{\Psi_0}\right)$.
Besides, $\ket{\phi_M(\theta)}$ is defined as
\begin{equation}
\ket{\phi_M(\theta)}:=U^{-1}_M\ket{S_M(\theta)},
\end{equation}
where $\ket{S_M(y)}$ is the state defined for the real number $y\in (0,1)$ as
\begin{equation}
\ket{S_M(y)}:=\frac{1}{\sqrt{M}}\sum_{x=0}^{M-1}e^{2\pi ixy}\ket{x},
\end{equation}
and $U^{-1}_M$ is the inverse of quantum Fourier transformation $U_M$ on $R_{\theta}$, that is,
\begin{equation}
U_M:\ket{x}\mapsto \frac{1}{\sqrt{M}}\sum_{y=0}^{M-1}e^{2\pi ixy/M}\ket{y}, x=0,1,...,M-1.
\end{equation}
We then measure $R_{\theta}$ in $\ket{\Phi_M(\theta)}$ and interpret the measurement outcome $\tilde{\theta}$ as a number in $[0,1)$ with $m$ fractional bits.
If $\tilde{\theta}>1/2$, we replace $\tilde{\theta}$ with $1-\tilde{\theta}$.
Then, this $\tilde{\theta}$ is close to $\theta$ with high probability:
\begin{eqnarray}
{\rm Pr}\left(\tilde{\theta}=\tilde{\theta}^\prime\right) &=& \frac{1}{2}\left[\left|\braket{\tilde{\theta}^\prime|\phi_M(\theta)}\right|^2+\left|\braket{1-\tilde{\theta}^\prime|\phi_M(1-\theta)}\right|^2\right] \nonumber \\
&=&\frac{\sin^2(M(\tilde{\theta}^\prime-\theta)\pi)}{M^2\sin^2((\tilde{\theta}^\prime-\theta)\pi)}\nonumber \\
&=:&G(\tilde{\theta}^\prime;\theta,M), \label{eq:G}
\end{eqnarray}
and this leads to
\begin{eqnarray}
{\rm Pr}\left(|\tilde{\theta}-\theta|<\frac{1}{M}\right)&=&\frac{\sin^2(M\delta\pi)}{M^2\sin^2(\delta\pi)}+\frac{\sin^2\left(M\left(\frac{1}{M}-\delta\right)\pi\right)}{M^2\sin^2\left(\left(\frac{1}{M}-\delta\right)\pi\right)} \nonumber \\
&\ge& \frac{8}{\pi^2}, \label{eq:QMCerr}
\end{eqnarray}
where $\delta=|\theta - \lfloor M\theta\rfloor/M|$.
Inequality (\ref{eq:QMCerr}) means that we can estimate $\theta$, or equivalently, $E_F$ with the worst-case error proportional to $M^{-1}$ by $O(M)$ calls of the integrand circuit $A$.
This is called the ``quadratic speedup" compared with classical Monte Carlo, where the error is proportional to the inverse square root of the number of calls to the integrand subroutine.

We here make some comments.
Firstly, it is sufficient to make only $S_0$ and $S_1$ controlled among the operations in $Q$ in order to make $Q$ controlled.
Two $A$'s do not have to be controlled.
We can easily see this as $Q$ becomes the identical transformation $I$ except an overall constant factor if $S_0$ and $S_1$ are replaced with $I$'s.
Therefore, if the integrand calculation included in $A$ makes the dominant contribution to complexity, making $Q$ be controlled increases complexity only slightly.
Secondly, in the QAE, the total number of integrand calculation and its inverse is nearly equal to $2M$, since the dominant contribution to the number comes from the about $M$ operations of controlled $Q$, and each of them contains one integrand calculation and one inverse.

\subsection{\label{sec:PRNMCReview}Use of pseudorandom number in the quantum algorithm for Monte Carlo integration}

We here briefly review the quantum method for Monte Carlo using PRNs\footnote{Of course, regardless of whether it is done in a classical or quantum way, Monte Carlo integration based on PRNs can induce additional errors, since PRNs are not truly random but deterministic. Every PRN generator does not have perfect statistical properties, for example, numbers in a PRN sequence inevitably have correlations to some extent. As far as the authors know, for PRN sequences which are widely used today, no established way to estimate errors in Monte Carlo integration due to statistical poorness is known. In many practical cases, we check randomness of a given PRN sequence through some statistical tests and use it neglecting errors if it passes the tests. The inventor of PCG claims that it passes TestU01\cite{TestU01}, a widely-used test suite for PRN.}, which is originally proposed in \cite{Miyamoto}.
When we apply the quantum algorithm for Monte Carlo to an extremely high-dimensional integration, it is necessary to generate as many RNs as the number of dimensions, in order to compute the integrand.
If we naively assign a register to each RN and create a superposition of possible values, the required qubit number increases in proportion to the number of dimensions.
In order to avoid this, we can adopt the following way.
First, as preparation, we choose a PRN sequence and set two registers, $R_{\rm samp}$ and $R_{\rm PRN}$.
Then, we create a superposition of integers, which specify the start points of the PRN sequence, on $R_{\rm samp}$.
For example, if we need $N_{\rm RN}$ RNs to compute the integrand, we can set the start points to the 1st, $(N_{\rm RN}+1)$-th, $(2N_{\rm RN}+1)$-th, ... elements in the sequence\footnote{Note that $N_{\rm RN}$ should be sufficiently smaller than the period of the PRN sequence. Conversely, we should choose a PRN sequence whose period is long enough.}. 
With each start point, we sequentially generate PRNs on $R_{\rm PRN}$.
This is possible because a PRN sequence is a deterministic sequence whose recursion equation is explicitly given, and in \cite{Miyamoto} we gave the implementation of one specific PRN generator, PCG, on quantum circuits.
Using the PRNs, we compute the integrand step by step.
Finally, the expectation value of the integrand is calculated by QAE.
In this way, since we need only $R_{\rm samp}$ and $R_{\rm PRN}$ to generate PRNs, the required qubit number is now independent of the number of dimensions and much smaller than the naive way.
The drawback is the increase of the circuit depth.

In this paper, we propose another way for Monte Carlo using PRNs, where we generate them not sequentially but in a quantum superposition, as explained in section \ref{sec:method}.

\section{\label{sec:method}The new method for Monte Carlo integration with speedup with respect to the number of dimension}

\begin{table*}[t]
	\caption{ The quantum registers used in the new method we propose.}
	\begin{tabular}{cm{40em}} \hline
		Symbol & Usage \\ \hline \hline
		$R_{\rm samp}$  & The register where we create the superposition of the indexes $j$ which specify one sample set of the stochastic variables $(\epsilon^{\rm PR}_{{\rm com},j},\epsilon^{\rm PR}_{1,j},...,\epsilon^{\rm PR}_{D,j})$. \\ \hline
		$R_{\rm dim}$ & The register where we create the superposition of the indexes $i$ which specify one individual stochastic variable $\epsilon^{\rm PR}_{i,j}$. \\ \hline
		$R_{\rm com}$ & The register where we output $\epsilon^{\rm PR}_{{\rm com},j}$.     \\ \hline
		$R_{\rm ind}$ & The register where we output $\epsilon^{\rm PR}_{i,j}$.     \\ \hline
		$R_{\vec{c}}$ & The register where we load $\vec{c}_{i}$.     \\ \hline
		$R_{f}$ & The register where we output $f(\epsilon^{\rm PR}_{{\rm com},j},\epsilon^{\rm PR}_{i,j};\vec{c}_i)$.     \\ \hline
		$R_{{\rm ph},f}$ & The single-qubit register where we encode $f(\epsilon^{\rm PR}_{{\rm com},j},\epsilon^{\rm PR}_{i,j};\vec{c}_i)$ as the amplitude of $\ket{1}$.     \\ \hline
		$R_{\rm ctr1}$ & The register which works as control bits in the inner QAE. After the inner QAE, the sum of  $f(\epsilon^{\rm PR}_{{\rm com},j},\epsilon^{\rm PR}_{i,j};\vec{c}_i)$'s over $i$ is encoded here.    \\ \hline
		$R_g$ & The register where we output $g\left(\sum_{i=1}^D f(\epsilon^{\rm PR}_{{\rm com},j},\epsilon^{\rm PR}_{i,j};\vec{c}_i)\right)$.     \\ \hline
		$R_{{\rm ph},g}$ & The single-qubit register where we encode $g\left(\sum_{i=1}^D f(\epsilon^{\rm PR}_{{\rm com},j},\epsilon^{\rm PR}_{i,j};\vec{c}_i)\right)$ as the amplitude of $\ket{1}$.     \\ \hline
		$R_{\rm ctr2}$ & The register which works as control bits in the outer QAE. After the outer QAE, $E_{\rm samp}$ is encoded here.    \\ \hline      
		\label{tbl:registers}
	\end{tabular}	
\end{table*}

In this section, we present a method to speed up the iterative calculation in computing the integrand in the quantum algorithm for Monte Carlo integration, which we call the {\it new method}.

\subsection{The problem}

First of all, let us clearly state the problem to which the new method can be applied.
Here and hereafter, we consider the Monte Carlo integration to calculate the expectation value $E\left[F\right]$ of the function $F$, which depends on some stochastic variables and takes the {\it separable} form given by
\begin{equation}
F(\epsilon_{\rm com},\{\epsilon_i\}_{i=1,...,D};\{\vec{c}_i\}_{i=1,...,D})=g\left(\sum_{i=1}^D f(\epsilon_{\rm com},\epsilon_i;\vec{c}_i)\right). \label{eq:FuncForm}
\end{equation}
That is, we can calculate $F$ by summing up the values of one common function $f$ with different inputs and operating the overall function $g$ to the sum.
Here, the meanings of the symbols are as follows.
$D$ is a natural number which satisfies $D\gg 1$.
$\epsilon_{\rm com}$ and $\epsilon_1,...,\epsilon_D$ are mutually independent stochastic variables.
The former is the {\it common stochastic variable}, which is used in all elements in the sum.
The others are {\it individual stochastic variables}.
They are independent and identically distributed and each of them is used in only one element in the sum. 
Although we hereafter consider $\epsilon_{\rm com}$ as a single stochastic variable for simplicity, it is straightforward to generalize the discussion to the case where $\epsilon_{\rm com}$ is a vector of multiple stochastic variables.
Totally, the number of the stochastic variables is $D+1$ and so is the dimension of the Monte Carlo integration. 
$\vec{c}_1,...,\vec{c}_D$ are sets of constant parameters.

\subsection{\label{sec:OutlineNew}The new method}

Then, let us consider how to calculate the expectation value $E[F]$ for the function $F$ in the form of (\ref{eq:FuncForm}).

The new method which we propose here is based on the PRN-approach of Monte Carlo integration on quantum computers, which we have explained in Section \ref{sec:PRNMCReview}.
In this approach, we sample many sets of the values of the stochastic variables using a PRN generator.
That is, in the current problem, we obtain the values of $\epsilon_{\rm com}$ and $\epsilon_1,...,\epsilon_D$ in the $j$-th sample set by sequentially applying the elements in a given PRN sequence $\{x_i\}_{i=1,2,...}$:
\begin{equation}
\epsilon^{\rm PR}_{{\rm com},j} := f_{\epsilon_{\rm com}}(x_{(j-1)(D+1)+1}), \epsilon^{\rm PR}_{i,j} := f_\epsilon(x_{(j-1)(D+1)+i+1}). \label{eq:svpr}
\end{equation}
Here, $f_{\epsilon_{\rm com}}$ and $f_\epsilon$ are the functions to transform the PRNs, which obey the uniform distribution in many cases, so that their distributions match that of $\epsilon_{\rm com}$ and $\epsilon_1,...,\epsilon_D$, respectively.
We will consider how to perform such transformations in section \ref{sec:eachpart}.
Then, $E[F]$ is estimated as
\begin{equation}
E_{\rm samp}:=\frac{1}{N_{\rm samp}}\sum_{j=1}^{N_{\rm samp}}g\left(\sum_{i=1}^D f(\epsilon^{\rm PR}_{{\rm com},j},\epsilon^{\rm PR}_{i,j};\vec{c}_i)\right), \label{eq:Esamp}
\end{equation}
where $N_{\rm samp}$ is the number of the samples.
The statistical error of the estimation $E_{\rm samp}$, that is, the confidence interval scales as $O(1/\sqrt{N_{\rm samp}})$.

The important point is that we can see $f(\epsilon^{\rm PR}_{{\rm com},j},\epsilon^{\rm PR}_{i,j};\vec{c}_i)$ as a function of $i$ and $j$.
That is, if we can implement the following circuits
\begin{itemize}
	\item $U_f$ \\
	This calculates $f(\epsilon_{\rm com},\epsilon_i;\vec{c}_i)$ for the given $\epsilon_{\rm com},\epsilon_i$ and $\vec{c}_i$.
	\item $U_{\epsilon_{\rm com}}$ \\
	This calculates $f_{\epsilon_{\rm com}}(x)$ for the given $x$.
	\item $U_{\epsilon}$\\
	This calculates $f_{\epsilon}(x)$ for the given $x$.
	\item $U_{J}$ \\
	This makes the PRN sequence $\{x_i\}_{i=1,2,...}$ jump to the given position.
	Here, we define {\it making $\{x_i\}_{i=1,2,...}$ jump} as the following operation: for a given integer $j\ge 1$, calculating $x_j$.
	\item $U_{\vec{c}}$ \\
	This loads $\vec{c}_i$ into a register for the given $i$.
\end{itemize}
we can implement the circuit to calculate $\epsilon^{\rm PR}_{{\rm com},j},\epsilon^{\rm PR}_{i,j}$ in (\ref{eq:svpr}) and then $f(\epsilon^{\rm PR}_{{\rm com},j},\epsilon^{\rm PR}_{i,j};\vec{c}_i)$ for the given $i$ and $j$.
Especially, availability of a formula for jump to a specified position is a beneficial feature of some kinds of PRNs, including PCG considered later, and it enables us to implement $U_J$ easily.
Including this point, we will explain how to implement these circuits in section {\ref{sec:eachpart}}.

If we can calculate the above function on a quantum computer, we can take the following way to calculate $E_{\rm samp}$ in (\ref{eq:Esamp}).
We call this a {\it nested QAE}, since it performs the summation over the sample index $j$ by QAE, which we call the {\it outer} QAE, and in each iteration in the outer QAE, another QAE, which we call the {\it inner} QAE, runs for the summation over $i$, the index of the terms.
The outline is as follows.
First, we make the superposition of states which correspond to the various sets of $(i,j)$.
Second, we calculate $f(\epsilon^{\rm PR}_{{\rm com},j},\epsilon^{\rm PR}_{i,j};\vec{c}_i)$ for the various pairs $(i,j)$ in quantum parallelism.
We then use the inner QAE: we sum up these values of $f$ over $i$ for each value of $j$ without sequential calculation and addition of $f$.
After operating $g$ on the sum, we use the outer QAE to get the sum over $j$, that is, $E_{\rm samp}$, avoiding sequential calculation again.

Note that the key factor is the map $(i,j) \mapsto f(\epsilon^{\rm PR}_{{\rm com},j},\epsilon^{\rm PR}_{i,j};\vec{c}_i)$.
Thanks to it, we can compute $f$ for various inputs in quantum parallelism and create the superposition of states corresponding to the various values of $f$, then finally apply the inner QAE to the superposition to estimate the sum of $f$'s with smaller complexity than sequential computation.
We again emphasize that using PRN enables us to implement this map.

The detailed steps of the new method are as follows.
Preparing the registers shown in Table \ref{tbl:registers}, each of which is initialized to $\ket{0}$, we perform the followings:
\begin{enumerate}
	\item Create $\frac{1}{\sqrt{N_{\rm samp}}}\sum_{j=1}^{N_{samp}}\ket{j}$ on $R_{\rm samp}$.
	\item With the input $j$ on $R_{\rm samp}$, calculate $\epsilon^{\rm PR}_{{\rm com},j}$ in (\ref{eq:svpr}) on $R_{\rm com}$ using $U_J$ and $U_{\epsilon_{\rm com}}$.
	\item Create $\frac{1}{\sqrt{D}}\sum_{i=1}^{D}\ket{i}$ on $R_{\rm dim}$.
	\item With the inputs $i$ on $R_{\rm dim}$ and $j$ on $R_{\rm samp}$, calculate $\epsilon^{\rm PR}_{i,j}$ in (\ref{eq:svpr}) on $R_{\rm ind}$ using $U_J$ and $U_{\epsilon}$.
	\item With the input $i$ on $R_{\rm dim}$, load $\vec{c}_{i}$ on $R_{\vec{c}}$ using $U_{\vec{c}}$.
	\item With the inputs on $R_{\rm com},R_\epsilon$ and $R_{\vec{c}}$, calculate $f(\epsilon^{\rm PR}_{{\rm com},j},\epsilon^{\rm PR}_{i,j};\vec{c}_i)$ on $R_f$.
	\item Using the rotation controlled by $R_f$, transform $R_{{\rm ph},f}$ to
	$\sqrt{1-f(\epsilon^{\rm PR}_{{\rm com},j},\epsilon^{\rm PR}_{i,j};\vec{c}_i)}\ket{0} + \sqrt{f(\epsilon^{\rm PR}_{{\rm com},j},\epsilon^{\rm PR}_{i,j};\vec{c}_i)}\ket{1}$.
	 Then, the probability that $R_{{\rm ph},f}$ is 1 under the condition that $R_{\rm samp}$ is $j$ is
	\begin{equation}
	S_j:=\frac{1}{D}\sum_{i=1}^D f(\epsilon^{\rm PR}_{{\rm com},j},\epsilon^{\rm PR}_{i,j};\vec{c}_i).
	\end{equation}
	\item Using the inner QAE, output $S_j$ on $R_{\rm ctr1}$. Strictly speaking, this step creates the state where the distribution of the value on $R_{\rm ctr1}$ is sharply peaked around $\theta_j$, which is defined through $\sin^2(\theta_j\pi):=S_j$ (see (\ref{eq:statetransf}) for the detail).
	\item With the input $\tilde{\theta}$ on $R_{\rm ctr1}$, calculate $\tilde{g}(\tilde{\theta}):=g(D\sin^2(\tilde{\theta}\pi))$, which is close to $g\left(\sum_{i=1}^D f(\epsilon^{\rm PR}_{{\rm com},j},\epsilon^{\rm PR}_{i,j};\vec{c}_i)\right)$ for $\tilde{\theta}\approx\theta_j$, on $R_g$.
	\item Using the rotation controlled by $R_g$, transform $R_{{\rm ph},g}$ to $\sqrt{1-\tilde{g}(\tilde{\theta})}\ket{0} + \sqrt{\tilde{g}(\tilde{\theta})}\ket{1}$.
	\item Using the outer QAE, estimate the probability of observing 1 on $R_{{\rm ph},g}$, which is nearly equal to $E_{\rm samp}$ (see (\ref{eq:p1})).
\end{enumerate}

The state is transformed through the above steps of 1-10  as follows.
Here, the first to tenth kets correspond to $R_{\rm samp},R_{\rm com},R_{\rm dim},R_{\rm ind},R_{\vec{c}},R_f,R_{{\rm ph},f},R_{{\rm ctr}1},R_g$ and $R_{{\rm ph},g}$, respectively.
\begin{widetext}
	\begin{eqnarray}
	& & \ket{0}\ket{0}\ket{0}\ket{0}\ket{0}\ket{0}\ket{0}\ket{0}\ket{0}\ket{0} \nonumber \\
	& \xrightarrow{1} & \frac{1}{\sqrt{N_{\rm samp}}}\sum_{j=1}^{N_{samp}}\ket{j}\ket{0}\ket{0}\ket{0}\ket{0}\ket{0}\ket{0}\ket{0}\ket{0}\ket{0} \nonumber \\
	& \xrightarrow{2} & \frac{1}{\sqrt{N_{\rm samp}}}\sum_{j=1}^{N_{samp}}\ket{j}\ket{\epsilon^{\rm PR}_{{\rm com},j}}\ket{0}\ket{0}\ket{0}\ket{0}\ket{0}\ket{0}\ket{0}\ket{0} \nonumber \\
	& \xrightarrow{3} & \frac{1}{\sqrt{N_{\rm samp}D}}\sum_{j=1}^{N_{samp}}\sum_{i=1}^{D}\ket{j}\ket{\epsilon^{\rm PR}_{{\rm com},j}}\ket{i}\ket{0}\ket{0}\ket{0}\ket{0}\ket{0}\ket{0}\ket{0} \nonumber \\
	& \xrightarrow{4,5} & \frac{1}{\sqrt{N_{\rm samp}D}}\sum_{j=1}^{N_{samp}}\sum_{i=1}^{D}\ket{j}\ket{\epsilon^{\rm PR}_{{\rm com},j}}\ket{i}\ket{\epsilon^{\rm PR}_{i,j}}\ket{\vec{c}_i}\ket{0}\ket{0}\ket{0}\ket{0}\ket{0} \nonumber \\
	& \xrightarrow{6} & \frac{1}{\sqrt{N_{\rm samp}D}}\sum_{j=1}^{N_{samp}}\sum_{i=1}^{D}\ket{j}\ket{\epsilon^{\rm PR}_{{\rm com},j}}\ket{i}\ket{\epsilon^{\rm PR}_{i,j}}\ket{\vec{c}_i}\ket{f(\epsilon^{\rm PR}_{{\rm com},j},\epsilon^{\rm PR}_{i,j};\vec{c}_i)}\ket{0}\ket{0}\ket{0}\ket{0} \nonumber \\
	& \xrightarrow{7} & \frac{1}{\sqrt{N_{\rm samp}D}}\sum_{j=1}^{N_{samp}}\sum_{i=1}^{D}\ket{j}\ket{\epsilon^{\rm PR}_{{\rm com},j}}\ket{i}\ket{\epsilon^{\rm PR}_{i,j}}\ket{\vec{c}_i}\ket{f(\epsilon^{\rm PR}_{{\rm com},j},\epsilon^{\rm PR}_{i,j};\vec{c}_i)}\left(\sqrt{1-f(\epsilon^{\rm PR}_{{\rm com},j},\epsilon^{\rm PR}_{i,j};\vec{c}_i)}\ket{0} + \sqrt{f(\epsilon^{\rm PR}_{{\rm com},j},\epsilon^{\rm PR}_{i,j};\vec{c}_i)}\ket{1}\right)\ket{0}\ket{0}\ket{0} \nonumber \\
	& =: &  \frac{1}{\sqrt{N_{\rm samp}}}\sum_{j=1}^{N_{samp}}\ket{j}\ket{\epsilon^{\rm PR}_{{\rm com},j}}\left(\sqrt{1-S_j}\ket{\Psi^{(j)}_0}+\sqrt{S_j}\ket{\Psi^{(j)}_1}\right)\ket{0}\ket{0}\ket{0}  \nonumber \\
	& \xrightarrow{8} & \frac{1}{\sqrt{2N_{\rm samp}}}\sum_{j=1}^{N_{samp}}\ket{j}\ket{\epsilon^{\rm PR}_{{\rm com},j}}\left(\ket{\Psi^{(j)}_+}\ket{\phi_M(\theta_j)}+\ket{\Psi^{(j)}_-}\ket{\phi_M(1-\theta_j)}\right)\ket{0}\ket{0} \nonumber \\
	&=& \frac{1}{\sqrt{2N_{\rm samp}}}\sum_{j=1}^{N_{samp}}\sum_{\tilde{\theta}\in I_M}\ket{j}\ket{\epsilon^{\rm PR}_{{\rm com},j}}\left(\braket{\tilde{\theta}|\phi_M(\theta_j)}\ket{\Psi^{(j)}_+}\ket{\tilde{\theta}}+\braket{1-\tilde{\theta}|\phi_M(1-\theta_j)}\ket{\Psi^{(j)}_-}\ket{1-\tilde{\theta}}\right)\ket{0}\ket{0} \nonumber \\
	& \xrightarrow{9} & \frac{1}{\sqrt{2N_{\rm samp}}}\sum_{j=1}^{N_{samp}}\sum_{\tilde{\theta}\in I_M}\ket{j}\ket{\epsilon^{\rm PR}_{{\rm com},j}}\left(\braket{\tilde{\theta}|\phi_M(\theta_j)}\ket{\Psi^{(j)}_+}\ket{\tilde{\theta}}+\braket{1-\tilde{\theta}|\phi_M(1-\theta_j)}\ket{\Psi^{(j)}_-}\ket{1-\tilde{\theta}}\right)\ket{g(D\sin^2(\tilde{\theta}\pi))}\ket{0} \nonumber \\
	& \xrightarrow{10} & \frac{1}{\sqrt{2N_{\rm samp}}}\sum_{j=1}^{N_{samp}}\sum_{\tilde{\theta}\in I_M}\ket{j}\ket{\epsilon^{\rm PR}_{{\rm com},j}}\left(\braket{\tilde{\theta}|\phi_M(\theta_j)}\ket{\Psi^{(j)}_+}\ket{\tilde{\theta}}+\braket{1-\tilde{\theta}|\phi_M(1-\theta_j)}\ket{\Psi^{(j)}_-}\ket{1-\tilde{\theta}}\right)\ket{g(D\sin^2(\tilde{\theta}\pi))}\left(\sqrt{1-\tilde{g}(\tilde{\theta})}\ket{0} + \sqrt{\tilde{g}(\tilde{\theta})}\ket{1}\right), \nonumber \\
	& & \label{eq:statetransf}
	\end{eqnarray}
where $I_M := \{0/M,1/M,...,(M-1)/M\}$, $M=2^{n_M}$, $n_M$ is the qubit number of $R_{\rm ctr1}$ and
		\begin{eqnarray}
		\ket{\Psi^{(j)}_0}&:=&\frac{1}{\sqrt{D}\sqrt{1-S_j}}\sum_{i=1}^{D}\sqrt{1-f(\epsilon^{\rm PR}_{{\rm com},j},\epsilon^{\rm PR}_{i,j};\vec{c}_i)}\ket{i}\ket{\epsilon^{\rm PR}_{i,j}}\ket{\vec{c}_i}\ket{f(\epsilon^{\rm PR}_{{\rm com},j},\epsilon^{\rm PR}_{i,j};\vec{c}_i)}\ket{0}, \nonumber \\
		\ket{\Psi^{(j)}_1}&:=&\frac{1}{\sqrt{D}\sqrt{S_j}}\sum_{i=1}^{D}\sqrt{f(\epsilon^{\rm PR}_{{\rm com},j},\epsilon^{\rm PR}_{i,j};\vec{c}_i)}\ket{i}\ket{\epsilon^{\rm PR}_{i,j}}\ket{\vec{c}_i}\ket{f(\epsilon^{\rm PR}_{{\rm com},j},\epsilon^{\rm PR}_{i,j};\vec{c}_i)}\ket{1}, \nonumber \\
		\ket{\Psi^{(j)}_{\pm}}&:=&\frac{1}{\sqrt{2}}\left(\ket{\Psi^{(j)}_1}\pm i\ket{\Psi^{(j)}_0}\right),
		\end{eqnarray}
\end{widetext}

\begin{table*}[t]
	\caption{ The quantum registers used in the previous method.}
	\begin{tabular}{cm{40em}} \hline
		Symbol & Usage \\ \hline \hline
		$R_{\rm samp}$  & The register where we create the superposition of the indexes $j$ which specify one sample set of the stochastic variables $(\epsilon^{\rm PR}_{{\rm com},j},\epsilon^{\rm PR}_{1,j},...,\epsilon^{\rm PR}_{D,j})$. \\ \hline
		$R_{\rm count}$ & The counter register which specifies a term in the integrand which we currently consider. \\ \hline
		$R_{\rm PRN}$ & The register where we sequentially generate PRNs.     \\ \hline
		$R_{\rm com}$ & The register where we output $\epsilon^{\rm PR}_{{\rm com},j}$.     \\ \hline
		$R_{\rm ind}$ & The register where we output $\epsilon^{\rm PR}_{i,j}$.     \\ \hline
		$R_{\vec{c}}$ & The register where we load $\vec{c}_{i}$.     \\ \hline
		$R_{f}$ & The register where we output $f(\epsilon^{\rm PR}_{{\rm com},j},\epsilon^{\rm PR}_{i,j};\vec{c}_i)$.     \\ \hline
		$R_{{\rm sum},f}$ & The register where we calculate the sum of $f$'s.     \\ \hline
		$R_g$ & The register where we output $g\left(\sum_{i=1}^D f(\epsilon^{\rm PR}_{{\rm com},j},\epsilon^{\rm PR}_{i,j};\vec{c}_i)\right)$.     \\ \hline
		$R_{{\rm ph},g}$ & The single-qubit register where we encode $g\left(\sum_{i=1}^D f(\epsilon^{\rm PR}_{{\rm com},j},\epsilon^{\rm PR}_{i,j};\vec{c}_i)\right)$ as the amplitude of $\ket{1}$.     \\ \hline
		$R_{\rm ctr}$ & The register which works as control bits in the QAE. After the QAE, $E_{\rm samp}$ is encoded here.    \\ \hline      
		\label{tbl:registersPrev}
	\end{tabular}	
\end{table*}

\noindent are the states in the tensor product space of $R_{\rm dim},R_{\rm ind},R_{\vec{c}},R_f$ and $R_{{\rm ph},f}$.
In (\ref{eq:statetransf}), we omit $R_{\rm ctr2}$ since it is used only in the step 11.
In the final state in (\ref{eq:statetransf}), the probability of observing 1 on $R_{{\rm ph},g}$ is
\begin{equation}
p_1 = \frac{1}{N_{\rm samp}}\sum_{j=1}^{N_{samp}}\sum_{\tilde{\theta}\in I_M}G(\tilde{\theta};\theta_j,M)\tilde{g}(\tilde{\theta}), \label{eq:p1}
\end{equation}
where $G$ is defined as (\ref{eq:G}).
Since $G(\tilde{\theta};\theta_j,M)$ has a sharp peak around $\tilde{\theta}=\theta_j$, $p_1$ is nearly equal to $E_{\rm samp}$.
We will discuss the error in section \ref{sec:error}.

\subsection{\label{sec:OutlinePrev}The previous method}

For completeness, we here outline the calculation procedure in which we use not the inner QAE but the simple iteration for the repeated calculation in the integrand.
Here and hereafter, we call this way the {\it previous method}.

In the previous method, we assume the availability of the circuit $U_P$, which progresses the PRN sequence $\{x_i\}_{i=1,2,...}$, in addition to the circuits listed in Section \ref{sec:OutlineNew}.
Here, {\it progressing $\{x_i\}_{i=1,2,...}$} is defined as the following operation: given the $i$-th element $x_i$, calculating the $(i+1)$-th element $x_{i+1}$.
Then, using $U_P$, we can sequentially generate PRNs as $\epsilon_1,\epsilon_2,...$ and calculate and sum up $f(\epsilon_{\rm com},\epsilon_i;\vec{c}_i)$ for $i=1,2,...$ step by step.
Concretely, preparing the registers in Table \ref{tbl:registersPrev}, each of which is initialized to $\ket{0}$, we perform the following procedure:

\begin{enumerate}
	\item Create $\frac{1}{\sqrt{N_{\rm samp}}}\sum_{j=1}^{N_{samp}}\ket{j}$ on $R_{\rm samp}$.
	\item With the input $j$ on $R_{\rm samp}$, calculate $x_{(j-1)(D+1)+1}$ on $R_{\rm PRN}$ by $U_J$, and using this, calculate $\epsilon^{\rm PR}_{{\rm com},j}$ in (\ref{eq:svpr}) on $R_{\rm com}$ by $U_{\epsilon_{\rm com}}$.
	\item Increment the value $i$ on $R_{\rm count}$ by 1.
	
	\item Using $U_P$, update $x_{(j-1)(D+1)+i}$ on $R_{\rm PRN}$ to $x_{(j-1)(D+1)+i+1}$ and using $U_{\epsilon}$, calculate $\epsilon^{\rm PR}_{i,j}$ in (\ref{eq:svpr}) on $R_{\rm ind}$.
	
	\item With the input $i$ on $R_{\rm count}$, load $\vec{c}_{i}$ on $R_{\vec{c}}$ using $U_{\vec{c}}$.
	
	\item With the inputs on $R_{\rm com},R_\epsilon$ and $R_{\vec{c}}$, calculate $f(\epsilon^{\rm PR}_{{\rm com},j},\epsilon^{\rm PR}_{i,j};\vec{c}_i)$ on $R_f$.
	\item Add the value on $R_f$ to $R_{{\rm sum},f}$.
	\item Uncompute $R_{\rm ind},R_{\vec{c}}$ and $R_f$.
	\item Repeat the steps 3 to 8 while the value $i$ on $R_{\rm count}$ satisfies $i\le D$ (in the $D$-th iteration, uncomputation in the step 8 in not necessary).
	When we stop, we have $DS_j=\sum_{i=1}^D f(\epsilon^{\rm PR}_{{\rm com},j},\epsilon^{\rm PR}_{i,j};\vec{c}_i)$ on $R_{{\rm sum},f}$, depending on the value $j$ on $R_{\rm samp}$. 
	\item With the input value on $R_{{\rm sum},f}$, calculate $g_j:=g\left(DS_j\right)$ on $R_g$.
	\item Using the rotation controlled by $R_g$, transform $R_{{\rm ph},g}$ to $\sqrt{1-g_j}\ket{0} + \sqrt{g_j}\ket{1}$.
	\item Using QAE, estimate the probability of observing 1 on $R_{{\rm ph},g}$, which is equal to $E_{\rm samp}$.
\end{enumerate}
Let us briefly comment on the difference between the new and previous methods.
First, the steps 3 to 8 in the new method, which correspond to the QAE-based calculation of $S_j$, are replaced by the steps 4 to 9, which are the $D$-time repetitions of progressing the PRN and calculation of $f$'s.
Besides, note that the previous method calculates $DS_j$ exactly, unlike the new method, which calculates $S_j$ with errors, that is, creates the superposition of values close to $S_j$.
Therefore, the probability of observing 1 on $R_{{\rm ph},g}$ in the previous method is exactly $E_{\rm samp}$, differently from the new method.

Through the steps 1 to 11, the state is transformed as follows.
Here, the first to tenth kets correspond to $R_{\rm samp},R_{\rm PRN},R_{\rm com},R_{\rm count},R_{\rm ind},R_{\vec{c}},R_f,R_{{\rm sum},f},R_g$ and $R_{{\rm ph},g}$, respectively.
\begin{widetext}
	\begin{eqnarray}
		& & \ket{0}\ket{0}\ket{0}\ket{0}\ket{0}\ket{0}\ket{0}\ket{0}\ket{0}\ket{0} \nonumber \\
		& \xrightarrow{1} & \frac{1}{\sqrt{N_{\rm samp}}}\sum_{j=1}^{N_{samp}}\ket{j}\ket{0}\ket{0}\ket{0}\ket{0}\ket{0}\ket{0}\ket{0}\ket{0}\ket{0} \nonumber \\
		& \xrightarrow{2} & \frac{1}{\sqrt{N_{\rm samp}}}\sum_{j=1}^{N_{samp}}\ket{j}\ket{x_{(j-1)(D+1)+1}}\ket{\epsilon^{\rm PR}_{{\rm com},j}}\ket{0}\ket{0}\ket{0}\ket{0}\ket{0}\ket{0}\ket{0} \nonumber \\
		& \xrightarrow{3} & \frac{1}{\sqrt{N_{\rm samp}}}\sum_{j=1}^{N_{samp}}\ket{j}\ket{x_{(j-1)(D+1)+1}}\ket{\epsilon^{\rm PR}_{{\rm com},j}}\ket{1}\ket{0}\ket{0}\ket{0}\ket{0}\ket{0}\ket{0} \nonumber \\
		& \xrightarrow{4} & \frac{1}{\sqrt{N_{\rm samp}}}\sum_{j=1}^{N_{samp}}\ket{j}\ket{x_{(j-1)(D+1)+2}}\ket{\epsilon^{\rm PR}_{{\rm com},j}}\ket{1}\ket{\epsilon^{\rm PR}_{1,j}}\ket{0}\ket{0}\ket{0}\ket{0}\ket{0} \nonumber \\
		& \xrightarrow{5} & \frac{1}{\sqrt{N_{\rm samp}}}\sum_{j=1}^{N_{samp}}\ket{j}\ket{x_{(j-1)(D+1)+2}}\ket{\epsilon^{\rm PR}_{{\rm com},j}}\ket{1}\ket{\epsilon^{\rm PR}_{1,j}}\ket{\vec{c}_1}\ket{0}\ket{0}\ket{0}\ket{0} \nonumber \\
		& \xrightarrow{6} & \frac{1}{\sqrt{N_{\rm samp}}}\sum_{j=1}^{N_{samp}}\ket{j}\ket{x_{(j-1)(D+1)+2}}\ket{\epsilon^{\rm PR}_{{\rm com},j}}\ket{1}\ket{\epsilon^{\rm PR}_{1,j}}\ket{\vec{c}_1}\ket{f(\epsilon^{\rm PR}_{{\rm com},j},\epsilon^{\rm PR}_{1,j};\vec{c}_1)}\ket{0}\ket{0}\ket{0} \nonumber \\
		& \xrightarrow{7} & \frac{1}{\sqrt{N_{\rm samp}}}\sum_{j=1}^{N_{samp}}\ket{j}\ket{x_{(j-1)(D+1)+2}}\ket{\epsilon^{\rm PR}_{{\rm com},j}}\ket{1}\ket{\epsilon^{\rm PR}_{1,j}}\ket{\vec{c}_1}\ket{f(\epsilon^{\rm PR}_{{\rm com},j},\epsilon^{\rm PR}_{1,j};\vec{c}_1)}\Ket{\sum_{i=1}^{1}f(\epsilon^{\rm PR}_{{\rm com},j},\epsilon^{\rm PR}_{1,j};\vec{c}_1)}\ket{0}\ket{0} \nonumber \\
		& \xrightarrow{8} & \frac{1}{\sqrt{N_{\rm samp}}}\sum_{j=1}^{N_{samp}}\ket{j}\ket{x_{(j-1)(D+1)+2}}\ket{\epsilon^{\rm PR}_{{\rm com},j}}\ket{1}\ket{0}\ket{0}\ket{0}\Ket{\sum_{i=1}^{1}f(\epsilon^{\rm PR}_{{\rm com},j},\epsilon^{\rm PR}_{1,j};\vec{c}_1)}\ket{0}\ket{0} \nonumber \\
		& \xrightarrow{9} & ... \nonumber \\
		& \xrightarrow{9} & \frac{1}{\sqrt{N_{\rm samp}}}\sum_{j=1}^{N_{samp}}\ket{j}\ket{x_{(j-1)(D+1)+D+1}}\ket{\epsilon^{\rm PR}_{{\rm com},j}}\ket{D}\ket{\epsilon^{\rm PR}_{D,j}}\ket{\vec{c}_D}\ket{f(\epsilon^{\rm PR}_{{\rm com},j},\epsilon^{\rm PR}_{D,j};\vec{c}_D)}\Bigg|\underbrace{\sum_{i=1}^{D}f(\epsilon^{\rm PR}_{{\rm com},j},\epsilon^{\rm PR}_{i,j};\vec{c}_i)}_{=DS_j}\Bigg>\ket{0}\ket{0} \nonumber \\
		& \xrightarrow{10} & \frac{1}{\sqrt{N_{\rm samp}}}\sum_{j=1}^{N_{samp}}\ket{j}\ket{x_{(j-1)(D+1)+D+1}}\ket{\epsilon^{\rm PR}_{{\rm com},j}}\ket{D}\ket{\epsilon^{\rm PR}_{D,j}}\ket{\vec{c}_D}\ket{f(\epsilon^{\rm PR}_{{\rm com},j},\epsilon^{\rm PR}_{D,j};\vec{c}_D)}\ket{DS_j}\ket{g(DS_j)}\ket{0} \nonumber \\
		& \xrightarrow{11} & \frac{1}{\sqrt{N_{\rm samp}}}\sum_{j=1}^{N_{samp}}\ket{j}\ket{x_{(j-1)(D+1)+D+1}}\ket{\epsilon^{\rm PR}_{{\rm com},j}}\ket{D}\ket{\epsilon^{\rm PR}_{D,j}}\ket{\vec{c}_D}\ket{f(\epsilon^{\rm PR}_{{\rm com},j},\epsilon^{\rm PR}_{D,j};\vec{c}_D)}\ket{DS_j}\ket{g(DS_j)}\left(\sqrt{1-g_j}\ket{0} + \sqrt{g_j}\ket{1}\right) \nonumber \\
		& & \label{eq:statetransfPrev}
	\end{eqnarray}
\end{widetext}

\subsection{\label{sec:eachpart}The parts of the circuit}

We here consider how to implement the component circuits listed in section \ref{sec:OutlineNew} and \ref{sec:OutlinePrev}.

\begin{itemize}
\item $U_f$ \\ \quad \\
This depends on the problems, so we here simply assume that it is implementable.
In section \ref{sec:example}, we consider its implementation for a concrete problem, that is, credit portfolio risk measurement.

\item $U_{\epsilon_{\rm com}},U_{\epsilon}$ \\ \quad \\
We here assume that PRNs obey the uniform distribution in $[0,1]$, as usual.
There are various ways to transform a uniform random number $x$ to a random number $y$ which obeys the desired distribution.
One is the inverse sampling method.
That is, we can transform $x$ as $y=\Phi^{-1}(x)$, where $\Phi^{-1}$ is the inverse of the cumulative distribution function (CDF) for the desired distribution.
In \cite{Kaneko}, the quantum circuit to calculate $\Phi^{-1}_{\rm SN}$, the inverse CDF for the standard normal distribution, is presented.
It is based on the piecewise polynomial approximation of $\Phi^{-1}_{\rm SN}$ presented in \cite{Hormann}.
We expect that the inverse CDFs for other distributions are also implemented in the similar way.

\item $U_P,U_{J}$  \\ \quad \\
Every PRN sequence has an explicit recursion formula.
Besides, for many widely-used PRN sequences, the simple formula to make the sequence jump to the desired position is explicitly given.
We can construct quantum circuits corresponding to these formulae.
Especially, in Section \ref{sec:example}, we will discuss how to construct, taking a concrete PRN generator, PCG, as an example.

\item $U_{\vec{c}}$  \\ \quad \\
If we can use a quantum random access memory (qRAM)\cite{Giovannetti}, we can implement $U_{\vec{c}}$ trivially.
Here, a qRAM is a quantum realization of associative data structure.
It refers to an index $i$ on a register and creates the state $\ket{d_i}$ which corresponds to the data $d_i$ associated with $i$ on another register.
That is, it performs the following operation: $\ket{i}\ket{0}\mapsto \ket{i}\ket{d_i}$.
Hereafter, we simply assume its availability.

\end{itemize}

\subsection{\label{sec:error}Error and complexity}

Sometimes, it is roughly said that in the QAE-based Monte Carlo method the number of repeated calculations of the integrand sufficient for the tolerance error $\delta$ is $\sim \delta^{-1}$.
On the basis of such a rough estimation, let us clarify the situation where the new method we propose is more advantageous than previous one.

In the current problem, calculation of $f(\epsilon^{\rm PR}_{{\rm com},j},\epsilon^{\rm PR}_{i,j};\vec{c}_i)$ is the most frequent procedure, so we focus on the number $N_f$ of this calculation and its relation to the error.
Here and hereafter, the word {\it calculation of $f$} means the repeated block in calculation of the sum of $f$'s and therefore includes some operations in addition to calculating $f$ itself.
More specifically, in the new method, calculation of $f$ corresponds to $Q$ in the inner QAE, or, in other words, the steps 3 to 7 in the calculation flow presented in Section \ref{sec:OutlineNew}.
On the other hand, in the previous method, calculation of $f$ corresponds to the steps 3 to 8 in the flow shown in Section \ref{sec:OutlinePrev}.

In the previous method, because we calculate the sum of $f$'s in (\ref{eq:Esamp}) by sequential calculations and additions of $f$ and use QAE only for the sum over the sample index $j$, it is necessary to take
\begin{equation}
N_{f,{\rm prev}}\sim D\delta^{-1}
\end{equation}
for the tolerance error $\delta$.
Here, the subscript `prev' means that the expression is for the previous method.
Note that the sequential evaluation of the sum of $f$'s causes no error. 
On the other hand, in the new method where the nested QAE is used, requiring that the error is at most $\delta$ in each QAE leads to
\begin{equation}
N_{f,{\rm new}}\sim \delta^{-2},
\end{equation}
where the subscript 'new' means that the expression is for the new method.
Therefore, comparing $N_{f,{\rm prev}}$ and $N_{f,{\rm new}}$, we see that the new method reduces $N_f$ if the inverse of the tolerance is smaller than the dimension of the integration, that is,
\begin{equation}
\delta^{-1} \lesssim D.
\end{equation}

The above estimation is illustrative but not strict since the result of the inner QAE is output as the superposition of the states, which correspond to the values distributing around the true value of the sum of $f$'s.
Let us evaluate the error by considering this distribution.
Here, we assume that $g$ is smooth, since in practical uses of Monte Carlo the integrand is at least piecewise smooth and a finite number of points where $g$ is non-smooth do not affect the integral.
Considering the fact that $G(\tilde{\theta};\theta_j,M)$ has a sharp peak around $\theta_j$, we approximate $\tilde{g}({\tilde{\theta}})=g(D\sin^2(\tilde{\theta}\pi))$ as the first degree Taylor expansion around $\tilde{\theta}=\theta_j$:
\begin{equation}
\tilde{g}(\tilde{\theta}) \simeq g(DS_j) + Dg^\prime(DS_j)(\sin^2(\tilde{\theta}\pi) - S_j),
\end{equation}
where we used $\sin^2(\theta_j\pi)=S_j$.
Using this and $\sum_{\tilde{\theta}\in I_M}G(\tilde{\theta};\theta_j,M)=1$, $p_1$ becomes
\begin{equation}
p_1 \simeq \frac{1}{N_{\rm samp}}\sum_{j=1}^{N_{\rm samp}}g(DS_j)\left(1 + \Delta(D,S_j,M)\right), \label{eq:p1approx}
\end{equation}
where the error term $\Delta(D,S_j,M)$ is defined as
\begin{equation}
\Delta(D,S_j,M) := \frac{Dg^\prime(DS_j)H(\theta_j,M)}{g(DS_j)}
\end{equation}
and $H$ is defined as
\begin{equation}
H(\theta_j,M):=\sum_{\tilde{\theta}\in I_M}G(\tilde{\theta};\theta_j,M)\left(\sin^2(\tilde{\theta}\pi) - \sin^2(\theta_j\pi)\right) \label{eq:H}.
\end{equation}
As explained in appendix \ref{sec:HEstim},
\begin{equation}
|H(\theta_j,M)|<\frac{1}{M} + O\left(\frac{1}{M^2}\right). \label{eq:HEstim}
\end{equation}
Therefore, the error $\Delta(D,S_j,M)$ is bounded as
\begin{equation}
|\Delta(D,S_j,M)| < \frac{DS_j|g^\prime(DS_j)|}{g(DS_j)}\frac{1/M}{S_j}+ O\left(\frac{1}{M^2}\right). \label{eq:Delta}
\end{equation}
(\ref{eq:Delta}) reasonably means the following.
In the usual situation where $DS_j|g^\prime(DS_j)|/g(DS_j)\sim 1$, which means that the change of the argument of $g$ by $O(1)$ factor leads to the change of $g$ by $O(1)$ factor, the deviation of $p_1$ from $E_{\rm samp}$ due to the inner QAE is negligible if $1/M$ is small compared with $S_j$.
$1/M\ll S_j$ can be rephrased that $R_{{\rm ctr},1}$, the output register for the inner QAE, has the large number of qubits enough to precisely estimate $\theta_j$, or equivalently, $S_j$.
In summary, it is required that
\begin{equation}
M>(l\delta_{\rm rel})^{-1},
\end{equation}
where $l$ is the typical scale of $S_j$, and $\delta_{\rm rel}$ is the tolerance relative error on $g(DS_j)$, and so the number $N_{f,{\rm QAE}1}$ of calculations of $f$ in the inner QAE, which is related to $M$ as $N_{f,{\rm QAE}1} \simeq M$, is at least $(l\delta_{\rm rel})^{-1}$.
Therefore, if
\begin{equation}
(l\delta_{\rm rel})^{-1} < D, \label{eq:condnewben}
\end{equation}
the new method reduces the number of calculations of $f$ by a factor
\begin{equation}
\frac{(l\delta_{\rm rel})^{-1}}{D}. \label{eq:reduc}
\end{equation}

We here make an important comment.
Although the new method can reduce the number of calculations of $f$, the total calculation time might not necessarily decrease.
This is because the steps in calculating $f$ are different between the previous and new methods.
In the sequential calculation of $f$ in the previous method, we progress the PRN sequence step by step.
On the other hand, in the new method, we make the PRN sequence jump to the specified position to get a RN input to $f$.
Usually, the jump takes a much larger computational cost than the progress.
If we write the times for {\it one} calculation of $f$ in the previous and new methods as $T_{\rm one,prev}$ and $T_{\rm one,new}$ respectively, the ratio of the total computational time in the new method to that in the previous method is
\begin{equation}
	\frac{T_{\rm one,new}}{T_{\rm one,prev}}\frac{(l\delta_{\rm rel})^{-1}}{D}. \label{eq:reducGate} 
\end{equation}

In section \ref{sec:example}, taking a concrete problem, credit portfolio risk measurement, and a concrete PRN generator, PCG, we will discuss the above point more rigorously and estimate the extent of computational time reduction by the new method.

\section{\label{sec:example}Example: credit portfolio risk measurement with PCG}

In this section, we consider credit portfolio risk measurement as an example problem where the new method can be applied, taking PCG\cite{PCG} as a concrete PRN generator.
First, we briefly explain the outlines of credit portfolio risk measurement and PCG, and then estimate the extent of complexity reduction.

\subsection{Credit portfolio risk measurement}

One of the representative problems to which Monte Carlo is often applied in finance is credit portfolio risk measurement.
Each bank has a credit portfolio, that is, a collection of many loans or debts, which is exposed to risks of defaults of obligors.
Banks evaluate such credit risks by some {\it risk measures}, which correspond to some kinds of estimation of the loss by defaults.
The major ones are the value-at-risk (VaR), the percentile point (say, 99\%) of loss distribution, and the conditional VaR (CVaR), the expectation value of loss under the condition that it exceeds the VaR.
Such quantities are usually calculated by some mathematical model, for example the Merton model\cite{Merton}, in combination with Monte Carlo.
The calculation in the Merton model with Monte Carlo on a quantum computer has already been considered in \cite{Egger,Miyamoto}.
For the details of the model and its implementation to a quantum computer, we here only refer to these papers.
The point we should note here is that this problem is actually in the scope of this paper.
That is, the integrand can be written as $g(L)$, where $L$ is the random loss and the function $g$ is set according to the type of the risk measure.
$L$ is calculated as
\begin{eqnarray}
L & = & \sum_{i=1}^{N_{\rm obl}} f(\epsilon_{\rm com}, \epsilon_i;E_i,\alpha_i,z_i) \nonumber \\
f(\epsilon_{\rm com}, \epsilon_i;E_i,\alpha_i,z_i) & = & E_i\Theta(Z_i,z_i) \nonumber \\
Z_i & = & \alpha_i \epsilon_{\rm com} + \sqrt{1-\alpha_i^2} \epsilon_i. \label{eq:loss}
\end{eqnarray}
Here, the meanings of the symbols are as follows.
$\Theta(x,y)$ is the indicator function, that is,
\begin{equation}
\Theta(x,y) =
\begin{cases}
1 ; x < y \\
0 ; {\rm otherwise}
\end{cases}.
\end{equation}
$N_{\rm obl}$ is the number of the obligors.
$E_i$ is the exposure of the $i$ th obligor.
Note that it must be normalized so that $E_i\le 1$.
For example, we may divide exposures by the largest one.
$\alpha_i,z_i$ are the model parameters for the $i$ th obligor; see \cite{Miyamoto} for the detail.
In addition to a common RN $\epsilon_{\rm com}$, we generate one RN $\epsilon_i$ for the $i$-th obligor to determine whether he defaults or not, which means the total number of RNs required to get one sample value of the loss is $N_{\rm obl}+1$.
For VaR, $g$ is taken as
\begin{equation}
g(L)=\Theta(L_\alpha,L),
\end{equation}
that is, we can search (e.g. binary search) $L_\alpha$ satisfying $E[g(L)]=\alpha$, which means $L_\alpha$ is the $(1-\alpha)$-percentile point of the loss.
For CVaR, we take
\begin{equation}
g(L)=CL\Theta(L_\alpha,L),
\end{equation}
where the VaR $L_\alpha$ is predetermined and $C$ is a normalization factor to make $g\le 1$.
As a whole, we can see that the integrand form matches (\ref{eq:FuncForm}).

\subsection{PCG}

Reference \cite{Miyamoto} picked up PCG\cite{PCG} as a PRN generator which can be implemented in a quantum circuit.
PCG is the combination of linear congruential generator (LCG) and permutation of bit string.
The $n$-th element of a PCG sequence $x_n$ is recursively defined as follows:
\begin{eqnarray}
\begin{cases}
\tilde{x}_{i+1} = (a\tilde{x}_{i}+c) \mod m & \\
x_i=f^{\rm perm}(\tilde{x}_{i}), &
\end{cases}
\label{eq:PCGProg}
\end{eqnarray}
where $\tilde{x}_{i}$ is the background LCG sequence, $a,c,m,\tilde{x}_0$ are integer parameters satisfying $a>0, c \geq 0, m>0,0\leq \tilde{x}_0 < m$ and $f^{\rm perm}$ is permutation of a bit string, for which \cite{PCG} presented some patterns.
Note that we can make LCG, and therefore PCG too, jump to the specified position by the following formula
\begin{equation}
\tilde{x}_{i} = \left(a^i\tilde{x}_{0}+\frac{c(a^i-1)}{a-1}\right) \mod m. \label{eq:PCGJump}
\end{equation}
For the further details of PCG, consult \cite{PCG}.

Thanks to the above jump formula, we can implement the jump operator $U_J$ for PCG.
In fact, we have already presented the circuit $U_J$ for such a jump in \cite{Miyamoto}, along with $U_P$ for the recursion formula (\ref{eq:PCGProg})\footnote{Note that, in \cite{Miyamoto}, $U_P$ and $U_J$ are represented by different symbols, $P_{\rm PRN}$ and $J_{\rm PRN}$, respectively.}.

\subsection{Reduction of complexity}

As discussed in section \ref{sec:error}, the new method reduces the number of calculations of $f$ if (\ref{eq:condnewben}) is satisfied.
However, the total complexity might not necessarily decrease, since the new method replaces the progress of the PRN sequence in the previous method with the jump, a more costly operation.

Considering this, let us estimate the extent of computational time reduction by the new method in credit portfolio risk measurement with PCG.
At first, we estimate complexity of one calculation of $f$, which is repeated most, in the two methods.
We here take T-count as a measure of computational time cost.
T-count is the number of T-gates used in a given quantum circuit.
Since the T-gate is expected to be most time-consuming in the {\it Clifford+T} gate set\cite{Zhou}, a widely-considered universal gate set, T-count is a widely-used metric of computational time cost.
Besides, we make the following assumptions on numbers of digits for various numbers:
\begin{itemize}
	\item We use the $n_{\rm PRN}$-bit PCG, and therefore so is the background LCG. We use only top $n_{\rm dig}$ digits for calculation, since lower bits have poorer statistical properties\cite{PCG}.
	\item For numerical numbers, we use $n_{\rm dig}$-bit fixed-point numbers.
\end{itemize} 
Then, we can estimate as follows.
\quad \\

\noindent
(1) the previous method

Among operations constituting calculation of $f$ in the previous method, it is sufficient to consider the following ones, which are more costly than others:
\begin{itemize}
	\item a progress of PCG
	
	As shown in (\ref{eq:PCGProg}), this consists of a modular multiplication, a modular addition and a permutation.
	As discussed in \cite{Kaneko}, the dominant contribution to complexity comes from a modular multiplication.
	If we perform this in the self-updating way in order to avoid adding qubits at every multiplication, we have to combine two non-self-updating modular multiplications into self-updating one.
	As a result, the T-count is $140n_{\rm PRN}^2$ as estimated in \cite{Kaneko}\footnote{In this paper, we take only the leading term for T-count, as in \cite{Kaneko}}.
	
	\item a RN conversion and uncomputation
	
	As mentioned above, we use the inverse sampling method combined with the piecewise polynomial approximation of $\Phi^{-1}_{\rm SN}$\cite{Hormann}.
	According to the estimation in \cite{Kaneko}, one conversion costs T-count of $105n_{\rm dig}^2 + 28n_{\rm dig}n_{\rm ICDF}$ and the total T-count is the double of it.
	Here, $n_{\rm ICDF}$ is the number of the intervals in the piecewise approximation.	
\end{itemize}
Calculation of $f$ makes only subdominant contributions to complexity, since it contains non-modular additions and multiplications, which is less costly than modular ones.
Similarly, increment of $R_{\rm count}$ and adding $f$ are also subdominant.
Besides, we assume that cost of loading/unloading $\vec{c}_i$ is subdominant.
Actually, a qRAM is designed so that only $O(n)$ quantum logic gates are activated while a record is loaded from a qRAM storing $2^n$ records\cite{Giovannetti}.
In the current case, loading parameters for an obligor requires activation of $O(n_{\rm obl})$ gates.

In total, T-count for a calculation of $f$ is
\begin{equation}
T_{\rm one,prev} \simeq 140n_{\rm PRN}^2+210n_{\rm dig}^2 + 56n_{\rm dig}n_{\rm ICDF}. \label{eq:Toneprev}
\end{equation}

\quad \\

\noindent
(2) the new method

In this case, calculation of $f$ is equivalent to $Q$ in the inner QAE.
Among the operations in it, the dominant contributors to complexity are the following:
\begin{itemize}
	\item two jumps of PCG
	
	Here, ``two" is because $Q$ contains $A$ and its inverse.
	As shown in (\ref{eq:PCGJump}), a jump contains a modular exponentiation and this makes the dominant contribution to complexity.
	A modular exponentiation can be constructed as $2n_{\rm exp}$ modular multiplications, where $n_{\rm exp}$ is the number of digit of the exponent\cite{Vedral}.
	From (\ref{eq:svpr}), we see that the exponent is now $(j-1)(N_{\rm obl}+1)+i+1$, since the dimension $D$ is now $N_{\rm obl}$.
	Here, $0\le i \le N_{\rm obl},0\le j \le N_{\rm samp}$.
	Therefore, the exponent can be expressed by $n_{\rm samp}+n_{\rm obl}$ bits, where for simplicity we assume that $N_{\rm samp}$ and $N_{\rm obl}$ are now powers of two: $N_{\rm samp}=2^{n_{\rm samp}},N_{\rm obl}=2^{n_{\rm obl}}$.
	As a result, T-count for a jump is that for a modular multiplication times $2(n_{\rm samp}+n_{\rm obl})$, that is, $140(n_{\rm samp}+n_{\rm obl})n_{\rm PRN}^2$.
	Two jumps cost doubly.
	
	\item two conversions of RN from uniform to standard normal
	
	Same as in the previous method.	
\end{itemize}
Other operations are subdominant for complexity:
\begin{itemize}
	\item Controlled $S_0$
	
	This is equivalent to a multiply-controlled Toffoli gate.
	It has T-count linear with respect to the number of the control qubits\cite{Selinger,Maslov}.
	
	\item Controlled $S_1$
	
	This is equivalent to just a controlled Z gate. 

	\item $f$ and loading/unloading $\vec{c}_i$

	Same as in the previous method.
	
	\item controlled rotation
	
	This has T-count linear with respect to the logarithm of the required accuracy\cite{Egger,Kliuchnikov,Amy}.
 
\end{itemize}

In total, T-count for a calculation of $f$ is
\begin{equation}
T_{\rm one,new} \simeq 280(n_{\rm samp}+n_{\rm obl})n_{\rm PRN}^2 + 210n_{\rm dig}^2+56n_{\rm dig}n_{\rm ICDF}. \label{eq:Tonenew}
\end{equation}

Then, let us take a typical setting in practical use and compare (\ref{eq:Toneprev}) and (\ref{eq:Tonenew}) in the setting. 
As typical values, we here set $n_{\rm dig}=16$, $n_{\rm PRN}=64$ \cite{PCG}, $n_{\rm ICDF}=109$ \cite{Hormann} and $n_{\rm samp}=n_{\rm obl}=20$, which correspond to $N_{\rm samp}=N_{\rm obl}=2^{20}\approx 10^6$.
For these values, (\ref{eq:Toneprev}) and (\ref{eq:Tonenew}) become
\begin{eqnarray}
T_{\rm one,prev} & \simeq & 7.2\times 10^5 \nonumber \\
T_{\rm one,new} & \simeq & 4.6\times 10^7,
\end{eqnarray}
respectively, and the ratio is 
\begin{equation}
\frac{T_{\rm one,new}}{T_{\rm one,prev}}=64. \label{eq:Tratio}
\end{equation}

Finally, we can compare the total T-counts in the whole processes of the previous and new methods. 
Combining (\ref{eq:reducGate}) and (\ref{eq:Tratio}), we obtain the ratio of the total T-counts as
\begin{equation}
	64 \frac{(l\delta_{\rm rel})^{-1}}{D}. \label{eq:TratioTot}
\end{equation}
This means that, if $(l\delta_{\rm rel})^{-1}/D$, the reduction ratio of the number of queries to calculations of $f$ by the new method is smaller than $1/64$, it is more beneficial than the previous one.
Then, let us consider a typical setting in credit portfolio risk measurement.
We assume that $l=10^{-2}$, which roughly corresponds to the situation where the total loss is 1\% of the total exposure, and $\delta_{\rm rel}=10^{-2}$.
Besides, we are now taking $D=N_{\rm obl}=2^{20}$.
These lead to the query number reduction ratio $(l\delta_{\rm rel})^{-1}/D\simeq 10^{-2}$, and finally the total T-count reduction ratio is about 0.64.
That is, we can reduce the total computational time by several tens of percent.

\section{\label{sec:summary}Summary}

In this paper, we present a version of the quantum method for Monte Carlo integration using PRNs.
The use of PRNs was originally proposed in \cite{Miyamoto} for the sake of reduction of qubits in extremely high-dimensional integrations such as credit portfolio risk measurement.
As an extension of this, the method proposed in this paper can reduce more complexity.
That is, in the case where the integrand has the separable form like (\ref{eq:FuncForm}), the method can reduce the number of repeated calculations over the separated terms compared with the previous method.
The key point is that if we use PRN, we can calculate $f$, a component of the integrand, as a function of the indices $i$ and $j$, which specify the RN and the sample respectively.
This makes it possible to compute $f$'s in quantum parallelism, not sequentially as in the previous method.
Combined with QAE, this leads to the reduction of the number of calculations of $f$, if the dimension (or equivalently the number of RNs) and the tolerance are large enough.
We should note that the new method can increase the time for one calculation of $f$ since it replaces the progress of the PRN sequence with the jump, which is more costly.
Therefore, the new method might not reduce the total computational time even if the query complexity decreases.
Nonetheless, taking T-count as a metric of computational time cost, we saw that the new method actually reduces the total T-count in a typical case of credit portfolio risk measurement with PCG, as shown in Sec. \ref{sec:example}

In the original proposal \cite{Miyamoto}, sequential computability of PRNs was the key feature to avoid generation of RNs on different registers and reduce qubits.
In this paper, another feature of PRN has been focused.
That is, since it is a deterministic sequence whose element can be calculated as a function of the index, we can compute PRNs in quantum parallelism and create a superposition of them.
In future works, we will explore the possibility to utilize such a feature in other ways and make quantum algorithm for Monte Carlo more efficient. 

\appendix

\section{\label{sec:HEstim}Proof of (\ref{eq:HEstim})}

We can transform $H(\theta_j,M)$ to
\begin{eqnarray}
& & H(\theta_j,M)= \nonumber \\
& & \qquad \frac{\sin^2(M\theta_j \pi)}{M^2}\left[M\cos(2\theta_j\pi) + \sin(2\theta_j\pi)\sum_{\tilde{\theta}\in I_M} \cot((\tilde{\theta}-\theta_j)\pi)\right]. \nonumber \\
& & \label{eq:HTransf}
\end{eqnarray}
Here, we used (\ref{eq:G}), some formulae on trigonometric functions and the relation $\sin^2(M(\tilde{\theta}-\theta_j)\pi)=\sin^2(M\theta_j \pi)$, which follows from $\tilde{\theta}\in I_M$.
We can show that the second term in the parenthesis in (\ref{eq:HTransf}) is of subleading order with respect to $1/M$, since the terms in the sum of cotangents over $\tilde{\theta}\in I_M$ nearly cancel out.
In the limit that $\theta_j \rightarrow \tilde{\theta}$ where $\tilde{\theta}\in I_M$, one term in the cotangent sum diverges but $H(\theta_j,M)$ itself goes to $0$ due to the overall factor $\sin^2(M\theta_j\pi)$.
As a result, we get
\begin{equation}
H(\theta_j,M)=\frac{\sin^2(M\theta_j\pi)\cos(2\theta_j\pi)}{M} + O\left(\frac{1}{M^2}\right).
\end{equation}
Since $|\sin^2(M\theta_j \pi)\cos(2\theta_j\pi)/M|<1/M$, we finally obtain (\ref{eq:HEstim}).

\end{document}